\documentclass[fleqn,twoside]{article}
\usepackage{espcrc2}
\usepackage{graphicx,epsfig,color}
\newcommand{\AmS}{{\protect\the\textfont2
  A\kern-.1667em\lower.5ex\hbox{M}\kern-.125emS}}
\title{Why the knee at 100 PeV could not be seen?
\thanks{Talk given for 22nd ECRS, Turku'2010}}

\author{Yu.V. Stenkin
\address{Institute for Nuclear Research of
Russian Academy of Sciences,\\ Moscow 117312, Russia}%
\thanks{email: (stenkin@sci.lebedev.ru)}}
\begin{document}

\begin{abstract}
It is shown that a "second (iron) knee" in cosmic ray spectrum
expected at energy about 100 PeV could not probably be found
there. The reason is very simple: the position of the "iron knee"
depends on an answer to the questions: "What do we see at 3-5 PeV?
Is the knee seen at this energy associated with proton or iron
primaries?"
\end{abstract}
\maketitle

\section{Introduction}

If the knee at 3-5 PeV in cosmic ray spectrum is a result of a
break in primary proton spectrum, as it was claimed by the KASCADE
collaboration, and energy of the knee for different primary
particles (Z, A) is proportional to charge Z (or atomic number A),
then one should expect "iron knee" existence at $\sim$100 (or 200)
PeV. Discovery of this "iron knee" was one of the main goals for
KASCADE-Grande experiment \cite{hau1}. But, the data obtained
in this experiment \cite{blu} as well as in all others, up to
date did not show any significant change of the spectrum slope in
the expected region.

On the other hand, as it is known, the conclusions of Tibet
AS$\gamma$ and Tibet-III experiments \cite{ame2,ame1} were the following: "Our
results shows that the main component responsible for the knee
structure of the all particle spectrum is heavier than helium
nuclei." It is in strong contradiction with the KASCADE
conclusion. If one believes in the Tibet experiment results then
one could expect an existence of the "proton knee" at energy lower
by a factor of 26 (or 56 ) than 3-5 PeV, i. e. close to 100 TeV.
Unfortunately, their energy threshold was put to 200 TeV and the
knee is seen only in comparison with all other data.

\section{Phenomenological approach}

The reason of the "puzzle" could be found if one takes into
account a new approach to the knee problem \cite{ste1,ste2}. In
this phenomenological approach primary spectrum could follow pure power
 law while the knee is expected to be seen
 in electromagnetic components and its position should occur
at energy of $\sim$100 TeV / nucleon for all primary particles. This
is a critical point in EAS development when the equilibrium
between hadronic and electromagnetic EAS components undergoes a
break at observation level. The primary masses A are spread from
A=1 for protons through A=56 for iron. Therefore, the knee visible
in PeV region should be connected with iron primaries while
"proton knee" is expected to be seen at $\sim$100 TeV. This
approach agrees with the Tibet AS$\gamma$  experiment conclusion
that the knee in PeV region is connected with heavy primaries. It
should be emphasized that the hybrid method used by the Tibet
experiment gave an advantage to this experiment in
comparison with a traditional EAS array, resulting in real primary
mass selection for individual events (at least for light
primaries).

\section{Experimental situation}

Compilation of the experimental data shown in \cite{blu,ame1}
makes us sure that there are no visible knee at energy $\sim$100
PeV. Small change of the slope (steepening then flattening and
then again steepening) shown by the KASCADE collaboration at the
European Cosmic Ray Symposium in Turku \cite{bert} was obtained by
only one method ($N_{e}$-$N_{\mu}$) while other methods gave
negative results (see fig. 1). There are no self-agreement of the
data and I think it should be proved by other experiments to be
regarded as an "iron knee").

But, a question arises: did anybody see the knee
at 100 TeV? The answer is "yes". A technical problem exists for
this region - this point coincides, as a rule, for the great bulk
of EAS arrays with their energy threshold. But, if the threshold
is put well below of 100 TeV then a "knee-like" behavior can be
seen. For example, such "knee-like" behavior can be found in
results of simulations (fig. 13 in \cite{ame1}).
\begin{figure}[bth!]
\begin{center}
\includegraphics[width=7cm]{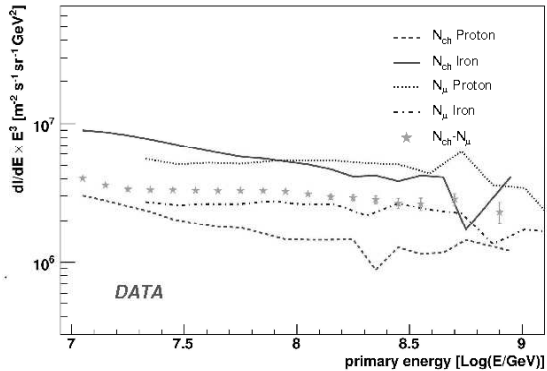}
\end{center}
\vspace{-4cm}
\caption{Energy spectra as recovered by KASCADE-Grande experiment using 3 %
different methods \cite{bert}}
\end{figure}
\begin{figure}[tbh!]
\begin{center}
\includegraphics[width=7.5cm]{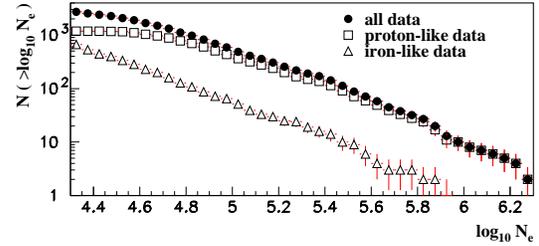}
\end{center}
\vspace{-1.5cm}
\caption{Integral spectra of the shower size $N_{e}$ obtained in
\cite{hau2}}
\end{figure}
Unfortunately their experimental spectrum starts just from 100 TeV
but, it is seen from fig.18 in \cite{ame2} that below this point
the spectrum should be flatter to agree with results of direct
spectrum measurements.
    Similar behavior of measured EAS size spectrum (fig. 2) can be found
even in early works of KASCADE (see fig. 13 in \cite{hau2} or
fig. 4 in \cite{hau3}). Moreover, the change of slope in the
EAS size spectrum at $N_{e}\approx 10^{4.8}$ is clearly caused by
the "proton-like events". Later this work has been forgotten. The
knee at $\sim$100 TeV (or $N_{e} \sim10^{5}$) was also observed in
the following experiments: HEGRA \cite{arq}, Carpet-2
\cite{ste3}, Grapes-III \cite{hay}, Tien-Shan \cite{nes}.

\section{Conclusion}

The problem of the cosmic ray knee in PeV region is still far from
its solution. The EAS technique elaborated many years ago allowed
physicists to investigate energy regions, which could not be
reached with direct cosmic ray measurements. But, as an indirect
method, it has many uncertainties and simplifications. Sometimes
the result depends on the suppositions made a priori. And the most
sensitive it is to a supposition on the existence of the knee in
primary cosmic ray spectrum.

On my opinion, traditional EAS arrays can not solve this very
complicated problem. One of the best array of the classical type -
KASCADE (KASCADE-GRANDE) - did not solve the problem of the knee.
It would be very difficult to make better classical array.  New
approaches and new ideas are needed.
    The PRISMA project proposed by us \cite{ste4}, would be an alternative
array aimed to the knee problem. It based on the idea that hadrons
form the EAS structure and thus hadrons should be the main EAS
component to be recorded and studied. A grid of a large number of
hadron sensitive scintillator detectors spread on the area of
$10^{4}$ - $10^{5}$ $m^{2}$ on the ground surface will be used to
record two main EAS component: hadronic (through thermal neutrons)
and electromagnetic. The project will have many advantages in
comparison with the traditional arrays: it will work as a huge
area hadronic calorimeter, it will have better core location
accuracy, better energy resolution, etc. It could give us a
possibility to measure EAS size spectra not only in electrons but
in hadrons and in muons as well. High altitude location is
preferable for such experiment. That is why the project could be
combined with other high altitude projects, such as LHAASO or
HAWC.

\section*{Acknowledgements} This work was supported by the RFBR
(grants 09-02-12380 and 08-02-01208).


\begin{thebibliography}{46}
\bibitem{hau1}
Haungs A., Apel W.D. et al. Investigating the second knee:
The KASCADE-GRANDE experiment. ArXiv: astro-ph/0508286, 2005.
\bibitem{blu}
Blumer V.J., Engel R. et al. Cosmic Rays from the Knee to the Highest Energies. ArXiv: 0904.0725v1, 2009.
\bibitem{ame2}
 Amenomori M. et al. Are protons still dominant at the knee of
the cosmic-ray energy spectrum?: ArXiv: Astro-ph/0511469v1, 2005.
\bibitem{ame1} Amenomori M. et al.: The All-particle spectrum of
primary cosmic rays in the wide energy range from $10^{14}$ to
$10^{17}$ eV observed with the Tiber-III air-shower array.:
Astrophys. J., v. 678, p. 1165, 2008. \bibitem{ste1} Stenkin
Yu.V.: "Does the "knee" in primary cosmic ray spectrum exist?":
Mod. Phys. Lett. A, 8(18), p. 1225-1234, 2003. \bibitem{ste2}
Stenkin Yu.V.: Phys. Atomic Nucl. V.71(1), p. 98, 2008.
\bibitem{bert} Bertaina M. et al. The cosmic ray energy spectrum
in the range $10^{16}$ - $10^{18}$ eV measured by KASCADE-Grande:
Report at 22 ECRS, Turku:
$http://ecrs2010.utu.fi/done/presentations/EDU2/2A_PA2_Wednesday/5_Bertaina.pdf$
\bibitem{hau2}Haungs A., Kempa J. et al. (KASCADE).: Report
FZKA6105, 1998. \bibitem{hau3}Haungs A., Kempa J. et al.
(KASCADE): Nucl. Phys. B, (Proc. Suppl.), 75A, p. 248.,1999.
\bibitem{arq} Arqueros F., Barrio J. A et al.
HEGRA-Collaboration.: Astron. and Astrophys. 359, p. 682-694,
2000. \bibitem{ste3} Stenkin Yu.V. et al.: Izvestia RAN, ser.
Fizich. V. 68 (11), p. 1611, 2004. \bibitem{hay} Hayashi Y. et
al.: Proc. 26th ICRC, v.1, p. 236, 1999. \bibitem{nes} Nesterova
.M., Chubenko A.P. et al.: Proc. Of 24th ICRC, Rome, v.2, p. 748,
1995. \bibitem{ste4} Stenkin Yu. V.: On the PRISMA project.:
ArXiv: 0902.0138v1 [Astro-ph.IM], 2009; Yu.V. Stenkin. Nucl. Phys.
B (Proc. Suppl.), v. 196, p. 293-296, 2009. \end{thebibliography}
\end{document}